# ANALYSIS OF ROTOR-BEARING SYSTEM USING THE TRANSFER MATRIX METHOD

**Mohammad T. Ahmadian**
Department of Mechanical Engineering
**Omid Ghasemalizadeh**
Department of Mechanical Engineering

**Hossein Sadeghi**
Department of Mechanical Engineering
**Mohammad Bonakdar**
Department of Mechanical Engineering

Center of Excellence in Dynamics, robotics, and Automation,
Sharif University of Technology, Tehran 11365-9567, Iran.

**ABSTRACT**
One of the methods to find the natural frequencies of rotating systems is the application of the transfer matrix method. In this method the rotor is modeled as several elements along the shaft which have their own mass and moment of inertia. Using these elements, the entire continuous system is discretized and the corresponding differential equation can be stated in matrix form. The bearings at the end of the shaft are modeled as equivalent spring and dampers which are applied as boundary conditions to the discretized system. In this paper the dynamics of a rotor-bearing system is analyzed, considering the gyroscopic effect. The thickness of the disk and bearings is also taken into account. Continuous model is used for shaft. Results Show that, the stiffness of the shaft and the natural frequencies of the system increase, while the amplitude of vibration decreases as a consequence of increasing the thickness of the bearing.

## 1 INTRODUCTION

Rotating shafts have an important role in power transmission and are used extensively in machines such as turbines, generators, electrical motors and etc. Various methods have been used in analyzing rotary systems. The Jeffcott model [1] for rotors assumes a massless shaft having a rigid rotor on it. The transfer matrix is a useful method for analyzing elastic structures [2]. The transfer matrices of many standard parts are given in handbooks. Sometimes, however, the required transfer matrix is not catalogued. It is often possible to find the transfer matrix by using simple dynamic equations or by using results that are tabulated in engineering handbooks. The merit of transfer matrix method lies in the fact that the dynamics of complicated systems can be simplified into a system of algebraic equations. Prohl [3] in 1945 suggested the transfer matrix method. lund and orcutt [4] in 1967 used this method with the continuous model for shaft and unbalance in disk taken into account. Kirk and Bansal [5] in 1975 used the transfer matrix method for calculating damped natural frequencies. Kang and Tan [6] in 1998 investigated discontinuity in shafts using this method. In this paper a Continuous model is used for the shaft and the gyroscopic effect is also taken into account. The key feature of this study is that the thickness of the disk is considered and the width of supporting end bearings is also modeled.

## 2 MODELING ASSUMPTIONS

In this analysis the vibration amplitude is assumed to be small. The Timoshenko beam theory is used for expressing the dynamics of the shaft. This theory assumes a constant transversal shear strain at the direction of thickness of the beam. To determine the shear forces, a correction factor (K) is used. Each section of the shaft is assumed to have two translational and two rotational motions about the perpendicular axes.

In this paper the fluid film journal bearing is considered for both ends of the shaft. Since the bearings have much more damping effect than the shaft, this effect is neglected for the shaft and the motion of shaft is assumed to be undamped. The bearings are modeled as equivalent spring and dampers which are applied as boundary conditions to the discretized system.

The transfer matrix of the shaft and the disk are obtained assuming that the gyroscopic effect is present and the disk has a finite thickness.

## 3 SHAFT TRANSFER MATRIX

An element of the deflected shaft in the yz and xz planes is depicted in figs.1 and 2, respectively. The subscripts $l$ and $r$ correspond to left and right sides of the element. The right-handed local coordinate system $(x_1, y_1, z_1)$ sticks to the shaft having its $z_1$ axis in the axial direction of the shaft. As shown below, the variation of shear forces and bending moments at the both ends of the element is calculated using partial derivatives.

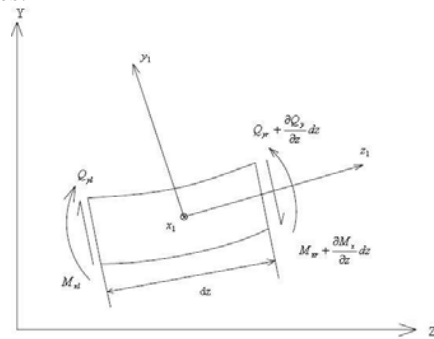

**Figure 1.** Element of shaft in zy plane

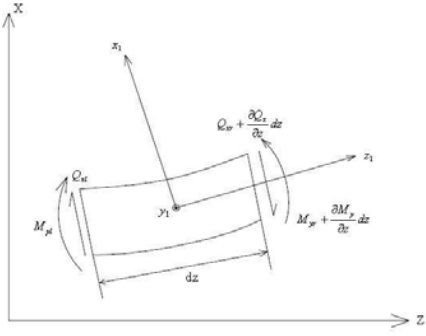

**Figure 2.** Element of shaft in zx plane

The resultant moment vectors for the element with unit length ($dz = 1$) is

$$\sum \mathbf{M} = \begin{bmatrix} -\dfrac{\partial M_x}{\partial z} + Q_y \\ \dfrac{\partial M_y}{\partial z} - Q_x \\ 0 \end{bmatrix} \quad (1)$$

and the angular momentum vector is

$$\vec{H} = \begin{bmatrix} -\rho I_x \dfrac{\partial \theta_x}{\partial t} \\ \rho I_y \dfrac{\partial \theta_y}{\partial t} \\ \rho I_z \omega \end{bmatrix} \quad (2)$$

Where $\theta_x$ and $\theta_y$ are the rotation of shaft about $x$ and $y$ axis, repectively and are assumed to be small as a consequence of assuming the vibration amplitude to be small, and $\omega$ is the spin velocity and $I_x$ and $I_y$ are the second moments of area which are assumed to be equal. Newton's second law for rotation is expressed as; [7]

$$\sum \mathbf{M} = \left(\dfrac{d\mathbf{H}}{dt}\right)_{xy} + \mathbf{\Omega} \times \mathbf{H} \quad (3)$$

where

$$\mathbf{\Omega} = \left\{ -\dfrac{\partial \theta_x}{\partial t} \quad \dfrac{\partial \theta_y}{\partial t} \quad \omega \right\}^T \quad (4)$$

is the angular velocity of the local coordinate system. Substituting equs. 1, 2 and 4 into equ. 3 results in

$$-\dfrac{\partial M_x}{\partial z} + Q_y = -\rho I_x \dfrac{\partial^2 \theta_x}{\partial t^2} + \rho \omega I_z \dfrac{\partial \theta_y}{\partial t} - \rho \omega I_y \dfrac{\partial \theta_y}{\partial t}$$

$$\dfrac{\partial M_y}{\partial z} - Q_x = \rho I_x \dfrac{\partial^2 \theta_y}{\partial t^2} + \rho \omega I_z \dfrac{\partial \theta_x}{\partial t} - \rho \omega I_x \dfrac{\partial \theta_x}{\partial t} \quad (5)$$

the resultant external force vector exerted on the element is

$$\sum \mathbf{F} = \begin{bmatrix} -\dfrac{\partial Q_x}{\partial z} \\ -\dfrac{\partial Q_y}{\partial z} \\ 0 \end{bmatrix} \quad (6)$$

Applying the Newton's second law to the element results in

$$-\dfrac{\partial Q_x}{\partial z} = \rho A \dfrac{\partial^2 x}{\partial t^2}, \quad -\dfrac{\partial Q_y}{\partial z} = \rho A \dfrac{\partial^2 y}{\partial t^2} \quad (7)$$

Equations 5 and 7 constitute the partial differential equation of the shaft. The following equations govern the bending moment and shear force in the shaft. [8]

$$M_x = EI_x \dfrac{\partial \theta_x}{\partial z}, \quad M_y = EI_x \dfrac{\partial \theta_y}{\partial z}$$

$$Q_x = KAG\left(\theta_y - \dfrac{\partial x}{\partial z}\right), \quad Q_y = KAG\left(\theta_x - \dfrac{\partial y}{\partial z}\right) \quad (8)$$

By replacing equ. 8 into equs. 5 and 7 the governing differential equations of the shaft in xy and yz planes are obtained, which simplifies to:

$$\dfrac{\partial^4 x}{\partial z^4} - \left(\dfrac{\rho}{KG} + \dfrac{\rho}{E}\right)\dfrac{\partial^4 x}{\partial z^2 \partial t^2} + \dfrac{\rho^2}{KEG}\dfrac{\partial^4 x}{\partial t^4}$$
$$+ \dfrac{\rho A}{EI}\dfrac{\partial^2 x}{\partial t^2} + \dfrac{2\rho\omega}{E}\left(\dfrac{\partial^3 y}{\partial z^2 \partial t} - \dfrac{\rho}{KG}\dfrac{\partial^3 y}{\partial t^3}\right) = 0$$

$$\dfrac{\partial^4 y}{\partial z^4} - \left(\dfrac{\rho}{KG} + \dfrac{\rho}{E}\right)\dfrac{\partial^4 y}{\partial z^2 \partial t^2} + \dfrac{\rho^2}{KEG}\dfrac{\partial^4 y}{\partial t^4}$$
$$+ \dfrac{\rho A}{EI}\dfrac{\partial^2 y}{\partial t^2} - \dfrac{2\rho\omega}{E}\left(\dfrac{\partial^3 x}{\partial z^2 \partial t} - \dfrac{\rho}{KG}\dfrac{\partial^3 x}{\partial t^3}\right) = 0 \quad (9)$$

The general solution to these PDEs is in the form of:

$$x = x_c(z)\cos\hat{\Omega}t + x_s(z)\sin\hat{\Omega}t$$
$$y = y_c(z)\cos\hat{\Omega}t + y_s(z)\sin\hat{\Omega}t \quad (10)$$

where $\hat{\Omega}$ is the natural frequency of the system. Substituting equ. 10 into equ. 9 results in four ODEs which require that the factors of *Sin* and *Cos* to be zero in order to have a nontrivial solution. The solution to the above set of ODE is in the form of:

$$x_c = u_c e^{\lambda z}, x_s = u_s e^{\lambda z}$$
$$y_c = v_c e^{\lambda z}, y_s = v_s e^{\lambda z} \quad (11)$$

By substituting the assumed solution into the four equations that have been created, a set of algebraic equations is obtained which is not stated here for the sake of brevity. In order to have a nontrivial solution the determinant of the coefficient matrix should vanish which results in the following characteristic equation:

$$\begin{pmatrix} -\rho A\Omega^2 KG + \rho^2\Omega^4 I_x + \rho\Omega^2 I_x \lambda^2 E + 2\rho^2\omega\Omega^3 I_x \\ +\rho\Omega^2 I_x \lambda^2 KG + \lambda^4 KGEI_x + 2\rho\omega\lambda^2\Omega KGI_x \end{pmatrix} \times$$
$$\begin{pmatrix} -\rho A\Omega^2 KG + \rho^2\Omega^4 I_x + \rho\Omega^2 I_x \lambda^2 E - 2\rho^2\omega\Omega^3 I_x \\ +\rho\Omega^2 I_x \lambda^2 KG + \lambda^4 KGEI_x - 2\rho\omega\lambda^2\Omega KGI_x \end{pmatrix} = 0 \quad (12)$$

The above equation has the following eight roots: $\pm\lambda_1, \pm i\lambda_2, \pm\lambda_3, \pm i\lambda_4$, which result the following statements for the four unknowns $(u_s, u_c, v_s, v_c)$

$$\begin{cases} u_s = v_c, & u_c = -v_s, \quad for\ (\pm i\lambda_2, \pm\lambda_1) \\ u_s = -v_c, & u_c = v_s, \quad for\ (\pm i\lambda_4, \pm\lambda_3) \end{cases} \quad (13)$$

Using the obtained roots, the displacement amplitudes may be expressed as;

$$\begin{aligned}
x_c &= b_1 \cosh(\lambda_1 z) + b_2 \sinh(\lambda_1 z) + b_3 \cos(\lambda_2 z) + b_4 \sin(\lambda_2 z) \\
&+ b_5 \cosh(\lambda_3 z) + b_6 \sinh(\lambda_3 z) + b_7 \cos(\lambda_4 z) + b_8 \sin(\lambda_4 z) \\
x_s &= b_9 \cosh(\lambda_1 z) + b_{10} \sinh(\lambda_1 z) + b_{11} \cos(\lambda_2 z) + b_{12} \sin(\lambda_2 z) \\
&+ b_{13} \cosh(\lambda_3 z) + b_{14} \sinh(\lambda_3 z) + b_{15} \cos(\lambda_4 z) + b_{16} \sin(\lambda_4 z) \\
y_c &= -b_1 \cosh(\lambda_1 z) - b_2 \sinh(\lambda_1 z) - b_3 \cos(\lambda_2 z) - b_4 \sin(\lambda_2 z) \\
&+ b_5 \cosh(\lambda_3 z) + b_6 \sinh(\lambda_3 z) + b_7 \cos(\lambda_4 z) + b_8 \sin(\lambda_4 z) \\
y_s &= b_9 \cosh(\lambda_1 z) + b_{10} \sinh(\lambda_1 z) + b_{11} \cos(\lambda_2 z) + b_{12} \sin(\lambda_2 z) \\
&- b_{13} \cosh(\lambda_3 z) - b_{14} \sinh(\lambda_3 z) - b_{15} \cos(\lambda_4 z) - b_{16} \sin(\lambda_4 z)
\end{aligned} \quad (14)$$

All the $b_i$ coefficients in the above equation are collected into the vector **B**.

$$\{\mathbf{B}\} = \{b_i\}_{16\times 1} \quad (15)$$

To obtain the transfer matrix of the shaft it is necessary to express the state vector in one side of the element in terms of the state vector in the other side. The state vector for the shaft is:

$$\begin{aligned}
\mathbf{S} = \{&x_c \quad x_s \quad y_c \quad y_s \quad \theta_{xc} \quad \theta_{xs} \quad \theta_{yc} \quad \theta_{ys} \\
& M_{xc} \quad M_{xs} \quad M_{yc} \quad M_{ys} \quad Q_{xc} \quad Q_{xs} \quad Q_{yc} \quad Q_{ys}\}^T
\end{aligned} \quad (16)$$

where s and c indices correspond to the *sin* and *cos* expressions as stated for x and y in equ. 10.
A vector of position derivatives is also defined as;

$$\begin{aligned}
\mathbf{W} = \{&x_c \quad x_c' \quad x_c'' \quad x_c''' \quad x_s \quad x_s' \quad x_s'' \quad x_s''' \\
& y_c \quad y_c' \quad y_c'' \quad y_c''' \quad y_s \quad y_s' \quad y_s'' \quad y_s'''\}^T
\end{aligned} \quad (17)$$

At two ends of the element, the above vector is equal to:

$$\begin{aligned} \{\mathbf{W}_{z=0}\} &= [\mathbf{A}_0]\{\mathbf{B}\} \\ \{\mathbf{W}_{z=L}\} &= [\mathbf{A}_L]\{\mathbf{B}\} \end{aligned} \quad (18)$$

where $[\mathbf{A}_0]$ and $[\mathbf{A}_L]$ are given in appendix 1.

By omitting the $\{\mathbf{B}\}$ vector in equs. 16 we conclude that;

$$\{\mathbf{W}_{z=L}\} = [\mathbf{A}_L][\mathbf{A}_0]^{-1}\{\mathbf{W}_{z=L}\} = [\mathbf{A}]\{\mathbf{W}_{z=0}\} \quad (19)$$

By expanding the components in the **W** vector, the relation between **W** and **S** may be found.

$$x = x, \quad x' = \theta_y - \frac{Q_x}{KAG}, \quad x'' = \frac{M_y}{EI_x} - \frac{1}{KAG}(-\rho A\ddot{x})$$

$$x''' = \frac{1}{EI_x}(Q_x + \rho I_x \ddot{\theta}_y + 2\rho I_x \omega \dot{\theta}_x) + \frac{\rho}{KG}(\ddot{\theta}_y - \frac{\ddot{Q}_x}{KAG})$$

$$y = y, \quad y' = \theta_x - \frac{Q_y}{KAG}, \quad y'' = \frac{M_x}{EI_x} - \frac{1}{KAG}(-\rho A\ddot{y}) \quad (20)$$

$$y''' = \frac{1}{EI_x}(Q_y + \rho I_x \ddot{\theta}_x - 2\rho I_x \omega \dot{\theta}_y) + \frac{\rho}{KG}(\ddot{\theta}_x - \frac{\ddot{Q}_y}{KAG})$$

By collecting the above relations into matrix form, the following relation is resulted

$$\{\mathbf{W}\} = [\mathbf{F}]\{\mathbf{S}\} \quad (21)$$

where $[\mathbf{F}]$ is given in appendix 1. Using equs. 19 and 21 the state vector in one side of the element may be related to that of the other side as follows:

$$\begin{aligned}
\{\mathbf{S}_{z=L}\} &= [\mathbf{F}]^{-1}\{\mathbf{W}_{z=L}\} = [\mathbf{F}]^{-1}[\mathbf{A}]\{\mathbf{W}_{z=0}\} \\
&= [\mathbf{F}]^{-1}[\mathbf{A}][\mathbf{F}]\{\mathbf{S}_{z=0}\} = [\mathbf{T}]\{\mathbf{S}_{z=0}\}
\end{aligned} \quad (22)$$

So the transfer matrix of the shaft is obtained from.

$$[\mathbf{T}]_{16\times 16} = [\mathbf{F}]^{-1}[\mathbf{A}_L][\mathbf{A}_0]^{-1}[\mathbf{F}] \quad (23)$$

## 4 DISK TRANSFER MATRIX

The disk free body diagram is depicted in Fig. 3.

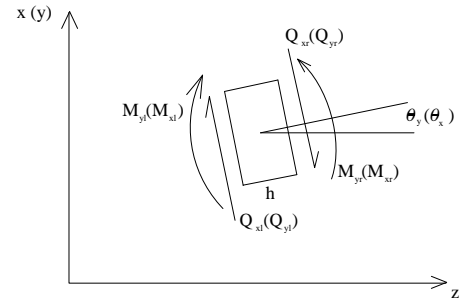

**Figure 3.** Free body diagram of disk

Here it is assumed that the disk is nonhomogeneous such that its center of mass is located by a small distance *r* off the geometric center. It is also assumed that *r* is such small that its effect on the inertia tensor may be neglected. So the inertia tensor of the disk is written as:

$$\mathbf{I} = \begin{bmatrix} I_d & 0 & 0 \\ 0 & I_d & 0 \\ 0 & 0 & 2I_d \end{bmatrix} \quad (24)$$

where $I_d$ is the mass moment of inertia of the disk about an axis passing through its center of gravity (CG). Again the Newton's second law is used to obtain the equations of motion for the disk. Application of Newton's second law to translational motion of the disk results in:

$$\begin{aligned}
m(\frac{\ddot{x}_l + \ddot{x}_r}{2} - r\omega^2 \cos\omega t) &= Q_{xl} - Q_{xr} \\
m(\frac{\ddot{y}_l + \ddot{y}_r}{2} - r\omega^2 \sin\omega t) &= Q_{xl} - Q_{xr}
\end{aligned} \quad (25)$$

Newton's second law in rotational motion is given in equ. 3. The angular momentum of disk is

$$\mathbf{H} = \mathbf{I} \times \mathbf{\Omega} \quad (26)$$

where $\mathbf{\Omega}$ is the same as than in equ. 4. The obtained angular momentum should be transformed into the global coordinate system using a rotation matrix $\mathbf{R}(\omega t)$.

$$\sum \mathbf{M} = \mathbf{R}(-\omega t)\dot{\mathbf{H}} + \mathbf{\Omega} \times (\mathbf{R}(-\omega t)\mathbf{H}) \quad (27)$$

The resultant of moment vectors acting on the disk is:

$$\sum \mathbf{M} = \begin{Bmatrix} M_{x\ell} - M_{xr} + \dfrac{h}{2} \times (Q_{y\ell} + Q_{yr}) \\ M_{yr} - M_{y\ell} + \dfrac{h}{2} \times (Q_{x\ell} + Q_{yr}) \\ 0 \end{Bmatrix} \quad (28)$$

Substituting equs. 26 and 28 into equ. 27 yields the equation of motion for rotation of the disk. Again by writing the degrees of freedom as in equ. 10, the above governing differential equations are simplified to some algebraic equations. Using the following additional equations;

$$\begin{aligned} \theta_{xr} &= \theta_{xl}, & \theta_{yr} &= \theta_{yl} \\ x_r &= x_l + h\theta_{xl}, & y_r &= y_l + h\theta_{yl} \end{aligned} \quad (29)$$

the state vector at both ends of the disk may now be related and the transfer matrix is obtained.

## 5  BEARING TRANSFER MATRIX

A linear model is assumed for bearings which uses two equivalent pairs of spring and dampers for each bearing as shown in fig. 4.

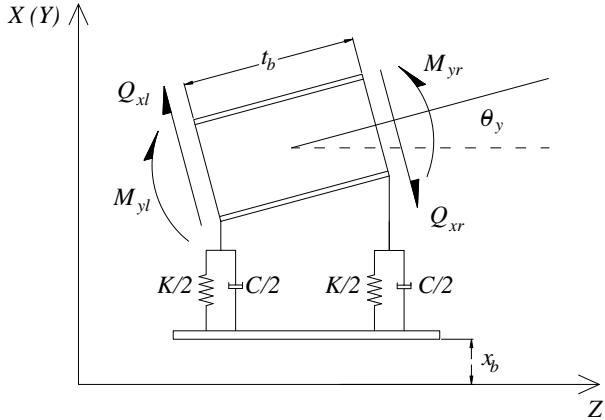

**Figure 4.** Bearing model

In this figure $t_b$ represents the length of the shaft that lies in the baring and $x_b$ or $y_b$ is the foundation displacement. This model assumes that each bearing can withstand moment in addition to transversal load. In a simplified model we assume that the bearing force vector can be written as;

$$\vec{F} = f(x,y)\hat{i} + g(x,y)\hat{j} \quad (30)$$

where

$$\begin{cases} f(x,y) = k_{xx}x + k_{xy}y \\ g(x,y) = k_{yx}x + k_{yy}y \end{cases} \quad (31)$$

the force and moment resultants are as follows;

$$\begin{aligned} \sum F_x &= Q_{xl} - Q_{xr} \\ &- \dfrac{k_{xx}}{2}(x_l - x_b) - \dfrac{k_{xy}}{2}(y_l - y_b) - \dfrac{k_{xx}}{2}(x_r - x_b) - \dfrac{k_{xy}}{2}(y_r - y_b) \\ &- \dfrac{c_{xx}}{2}(\dot{x}_l - \dot{x}_b) - \dfrac{c_{xy}}{2}(\dot{y}_l - \dot{y}_b) - \dfrac{c_{xx}}{2}(\dot{x}_r - \dot{x}_b) - \dfrac{c_{xy}}{2}(\dot{y}_r - \dot{y}_b) \\[4pt]
\sum F_y &= Q_{yl} - Q_{yr} \\
&- \dfrac{k_{yy}}{2}(y_l - y_b) - \dfrac{k_{yx}}{2}(x_l - x_b) - \dfrac{k_{yy}}{2}(y_r - y_b) - \dfrac{k_{yx}}{2}(x_r - x_b) \\
&- \dfrac{c_{yy}}{2}(\dot{y}_l - \dot{y}_b) - \dfrac{c_{xy}}{2}(\dot{x}_l - \dot{x}_b) - \dfrac{c_{yy}}{2}(\dot{y}_r - \dot{y}_b) - \dfrac{c_{yx}}{2}(\dot{x}_r - \dot{x}_b) \\[4pt]
\sum M_y &= M_{yr} - M_{yl} + \dfrac{k_{xx}}{4}(x_l - x_b)t_b \\
&- \dfrac{t_b}{2}(Q_{xl} - Q_{xr}) + \dfrac{k_{xy}}{4}(y_l - y_b)t_b - \dfrac{k_{xx}}{4}(x_r - x_b)t_b \\
&- \dfrac{k_{xy}}{4}(y_r - y_b)t_b + \dfrac{c_{xx}}{4}(\dot{x}_l - \dot{x}_b)t_b + \dfrac{c_{xy}}{4}(\dot{y}_l - \dot{y}_b)t_b \\
&- \dfrac{c_{xx}}{4}(\dot{x}_r - \dot{x}_b)t_b - \dfrac{c_{xy}}{4}(\dot{y}_r - \dot{y}_b)t_b \\[4pt]
\sum M_x &= M_{xl} - M_{xr} - \dfrac{k_{yy}}{4}(y_l - y_b)t_b \\
&+ \dfrac{t_b}{2}(Q_{yl} - Q_{yr}) - \dfrac{k_{yx}}{4}(x_l - x_b)t_b + \dfrac{k_{yy}}{4}(y_r - y_b)t_b \\
&+ \dfrac{k_{yx}}{4}(x_r - x_b)t_b - \dfrac{c_{yy}}{4}(\dot{y}_l - \dot{y}_b)t_b - \dfrac{c_{yx}}{4}(\dot{x}_l - \dot{x}_b)t_b \\
&+ \dfrac{c_{yy}}{4}(\dot{y}_r - \dot{y}_b)t_b + \dfrac{c_{yx}}{4}(\dot{x}_r - \dot{x}_b)t_b \end{aligned} \quad (32)$$

Using the Newton's second law in rotation and translation and by using the following geometrical additional equations;

$$\begin{aligned} \theta_{xr} &= \theta_{xl} = \theta_x, & \theta_{yr} &= \theta_{yl} = \theta_y \\ x_r &= x_l + t_b\theta_y, & y_r &= y_l + t_b\theta_x \end{aligned} \quad (33)$$

the governing differential equation of the bearing is obtained. Again by assuming harmonic motion for all degrees of freedom (as equ. 10), the governing algebraic equation is obtained which is used to calculate the transfer matrix of the bearing.

To obtain the transfer matrix between two desired points of the system, the transfer matrices of the individual elements located between the points should be multiplied consequently. If the transfer matrix of the whole system is available, applying the boundary conditions on the state vector of both ends will result in a relation among the unknown properties of the system.

## 6  RESULTS

To investigate the applicability of this method in dynamic analysis of rotor-bearing systems, an example is provided. A shaft carrying several disks is supported on two bearings on both ends as shown in fig. 5.

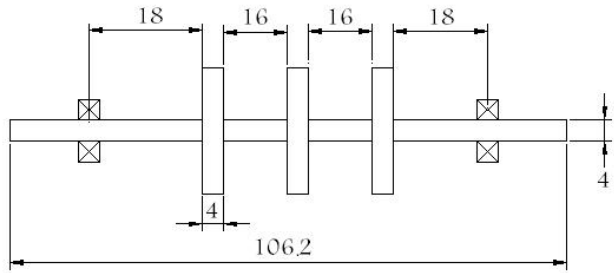

**Figure 5.** Schematic figure of a rotor

In this example a flexible shaft is carrying three similar disks and two bearings are supporting the shaft. The parameters of the system are given in table 1.

Table 1. Parameters of the rotor-bearing system.

| Modulus of elasticity $(N/cm^2)$ | $2.07 \times 10^7$ |
|---|---|
| Shear modulus $(N/cm^2)$ | $1.29 \times 10^7$ |
| Density $(kg/cm^3)$ | $7.75 \times 10^{-3}$ |
| Disc mass $(kg)$ | 13.47 |
| Polar moment of inertia $(kg.cm^2)$ | 1020 |
| Mass moment of inertia $(kg.cm)$ | 512 |
| Unbalance of disc $(kg.cm)$ | 0.01347 |
| Shaft diameter $(cm)$ | 4 |

When the disk width is ignored the amplitude-rotational speed curve in fig. 6 is obtained.

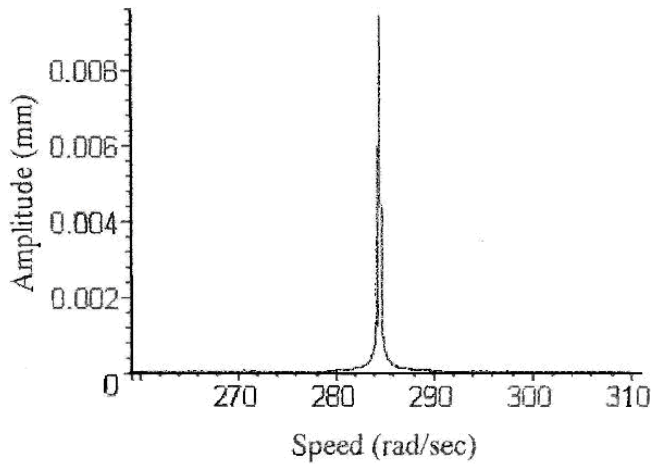

**Figure 5.** response of the system when ignoring the disk width

In the above figure the pick amplitude corresponds to the first natural frequency of the system which agrees with approximate methods available such as the Dunkerley's method. When the thickness of the disk is also taken into account, the response curve changes significantly as depicted in fig. 6.

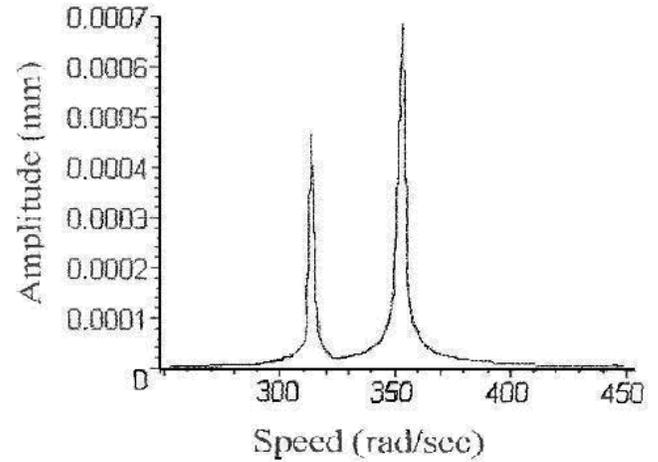

**Figure 6.** response of the system when considering the disk width

It is shown that the first natural frequency of the system has increased compared to the previous case. The increase of the natural frequency is attributed to the fact that by considering the disk width, the effective length of the shaft decreases and hence the system becomes more rigid. The decrease in the amplitude of vibration is another consequence of thickening the disk.

The real case can be achieved when both the disk and bearing thicknesses are taken into account. Figure 7 corresponds to this case.

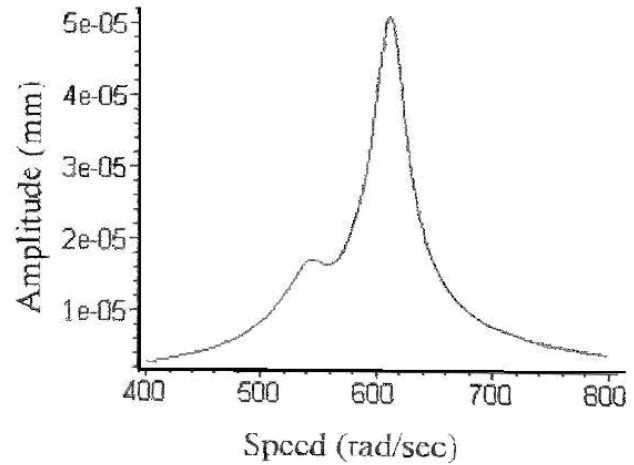

**Figure 7.** response of the system when considering both the disk and bearing widths

It is observed from fig. 7 that by considering the width of bearing the natural frequency increases again while the amplitude of vibration has decreased.

## 7 CONCLUSIONS

In this paper the matrix transfer method is applied to rotor-bearing system. The transfer matrix of individual elements including the shaft, disk and the bearing are obtained. The merit of this approach over the conventional approximate methods is that the thickness of the disks and the bearings may also be considered. It is shown that the thickness of disks and bearing have a considerable effect on the vibration behavior of the system. This method can be applied to rotor-bearing systems with various configurations.

## APPENDIX 1- SOME TRANSFER MATRICES

$$A_0 = \begin{bmatrix} 1 & 0 & 1 & 0 & 1 & 0 & 1 & 0 & 0 & 0 & 0 & 0 & 0 & 0 & 0 & 0 \\ 0 & \lambda_1 & 0 & \lambda_2 & 0 & \lambda_3 & 0 & \lambda_4 & 0 & 0 & 0 & 0 & 0 & 0 & 0 & 0 \\ \lambda_1^2 & 0 & -\lambda_2^2 & 0 & \lambda_3^2 & 0 & -\lambda_4^2 & 0 & 0 & 0 & 0 & 0 & 0 & 0 & 0 & 0 \\ 0 & \lambda_1^3 & 0 & -\lambda_2^3 & 0 & \lambda_3^3 & 0 & -\lambda_4^3 & 0 & 0 & 0 & 0 & 0 & 0 & 0 & 0 \\ 0 & 0 & 0 & 0 & 0 & 0 & 0 & 0 & 1 & 0 & 1 & 0 & 1 & 0 & 1 & 0 \\ 0 & 0 & 0 & 0 & 0 & 0 & 0 & 0 & 0 & \lambda_1 & 0 & \lambda_2 & 0 & \lambda_3 & 0 & \lambda_4 \\ 0 & 0 & 0 & 0 & 0 & 0 & 0 & 0 & \lambda_1^2 & 0 & -\lambda_2^2 & 0 & \lambda_3^2 & 0 & -\lambda_4^2 & 0 \\ 0 & 0 & 0 & 0 & 0 & 0 & 0 & 0 & 0 & \lambda_1^3 & 0 & -\lambda_2^3 & 0 & \lambda_3^3 & 0 & -\lambda_4^3 \\ 0 & 0 & 0 & 0 & 0 & 0 & 0 & 0 & 1 & 0 & 1 & 0 & -1 & 0 & -1 & 0 \\ 0 & 0 & 0 & 0 & 0 & 0 & 0 & 0 & 0 & \lambda_1 & 0 & \lambda_2 & 0 & -\lambda_3 & 0 & -\lambda_4 \\ 0 & 0 & 0 & 0 & 0 & 0 & 0 & 0 & \lambda_1^2 & 0 & -\lambda_2^2 & 0 & -\lambda_3^2 & 0 & \lambda_4^2 & 0 \\ 0 & 0 & 0 & 0 & 0 & 0 & 0 & 0 & 0 & \lambda_1^3 & 0 & -\lambda_2^3 & 0 & -\lambda_3^3 & 0 & \lambda_4^3 \\ -1 & 0 & -1 & 0 & 1 & 0 & 1 & 0 & 0 & 0 & 0 & 0 & 0 & 0 & 0 & 0 \\ 0 & -\lambda_1 & 0 & -\lambda_2 & 0 & \lambda_3 & 0 & \lambda_4 & 0 & 0 & 0 & 0 & 0 & 0 & 0 & 0 \\ -\lambda_1^2 & 0 & \lambda_2^2 & 0 & \lambda_3^2 & 0 & -\lambda_4^2 & 0 & 0 & 0 & 0 & 0 & 0 & 0 & 0 & 0 \\ 0 & -\lambda_1^3 & 0 & \lambda_2^3 & 0 & \lambda_3^3 & 0 & -\lambda_4^3 & 0 & 0 & 0 & 0 & 0 & 0 & 0 & 0 \end{bmatrix}$$

$$A_L[1...8,1...8] = \begin{bmatrix} \cosh(\lambda_1 L) & \sinh(\lambda_1 L) & \cos(\lambda_2 L) & \sin(\lambda_2 L) & \cosh(\lambda_3 L) & \sinh(\lambda_3 L) & \cos(\lambda_4 L) & \sin(\lambda_4 L) \\ \sinh(\lambda_1 L)\lambda_1 & \cosh(\lambda_1 L)\lambda_1 & -\sin(\lambda_2 L)\lambda_2 & \cos(\lambda_2 L)\lambda_2 & \sinh(\lambda_3 L)\lambda_3 & \cosh(\lambda_3 L)\lambda_3 & -\sin(\lambda_4 L)\lambda_4 & \cos(\lambda_4 L)\lambda_4 \\ \cosh(\lambda_1 L)\lambda_1^2 & \sinh(\lambda_1 L)\lambda_1^2 & -\cos(\lambda_2 L)\lambda_2^2 & -\sin(\lambda_2 L)\lambda_2^2 & \cosh(\lambda_3 L)\lambda_3^2 & \sinh(\lambda_3 L)\lambda_3^2 & -\cos(\lambda_4 L)\lambda_4^2 & -\sin(\lambda_4 L)\lambda_4^2 \\ \sinh(\lambda_1 L)\lambda_1^3 & \cosh(\lambda_1 L)\lambda_1^3 & \sinh(\lambda_2 L)\lambda_2^3 & -\cos(\lambda_2 L)\lambda_2^3 & \sinh(\lambda_3 L)\lambda_3^3 & \cosh(\lambda_3 L)\lambda_3^3 & \sinh(\lambda_4 L)\lambda_4^3 & -\cos(\lambda_4 L)\lambda_4^3 \\ 0 & 0 & 0 & 0 & 0 & 0 & 0 & 0 \\ 0 & 0 & 0 & 0 & 0 & 0 & 0 & 0 \\ 0 & 0 & 0 & 0 & 0 & 0 & 0 & 0 \\ 0 & 0 & 0 & 0 & 0 & 0 & 0 & 0 \end{bmatrix}$$

$A_L[1...8, 9...16] =$

$$\begin{bmatrix} 0 & 0 & 0 & 0 & 0 & 0 & 0 & 0 \\ 0 & 0 & 0 & 0 & 0 & 0 & 0 & 0 \\ 0 & 0 & 0 & 0 & 0 & 0 & 0 & 0 \\ 0 & 0 & 0 & 0 & 0 & 0 & 0 & 0 \\ -\cosh(\lambda_1 L) & -\sinh(\lambda_1 L) & -\cos(\lambda_2 L) & -\sin(\lambda_2 L) & \cosh(\lambda_3 L) & \sinh(\lambda_3 L) & \cos(\lambda_4 L) & \sin(\lambda_4 L) \\ -\sinh(\lambda_1 L)\lambda_1 & -\cosh(\lambda_1 L)\lambda_1 & \sin(\lambda_2 L)\lambda_2 & -\cos(\lambda_2 L)\lambda_2 & \sinh(\lambda_3 L)\lambda_3 & \cosh(\lambda_3 L)\lambda_3 & -\sin(\lambda_4 L)\lambda_4 & \cos(\lambda_4 L)\lambda_4 \\ -\cosh(\lambda_1 L)\lambda_1^2 & -\sinh(\lambda_1 L)\lambda_1^2 & \cos(\lambda_2 L)\lambda_2^2 & \sin(\lambda_2 L)\lambda_2^2 & \cosh(\lambda_3 L)\lambda_3^2 & \sinh(\lambda_3 L)\lambda_3^2 & -\cos(\lambda_4 L)\lambda_4^2 & -\sin(\lambda_4 L)\lambda_4^2 \\ -\sinh(\lambda_1 L)\lambda_1^3 & -\cosh(\lambda_1 L)\lambda_1^3 & -\sin(\lambda_2 L)\lambda_2^3 & \cos(\lambda_2 L)\lambda_2^3 & \sinh(\lambda_3 L)\lambda_3^3 & \cosh(\lambda_3 L)\lambda_3^3 & \sin(\lambda_4 L)\lambda_4^3 & -\cos(\lambda_4 L)\lambda_4^3 \end{bmatrix}$$

$A_L[9...16, 1...8] =$

$$\begin{bmatrix} 0 & 0 & 0 & 0 & 0 & 0 & 0 & 0 \\ 0 & 0 & 0 & 0 & 0 & 0 & 0 & 0 \\ 0 & 0 & 0 & 0 & 0 & 0 & 0 & 0 \\ 0 & 0 & 0 & 0 & 0 & 0 & 0 & 0 \\ \cosh(\lambda_1 L) & \sinh(\lambda_1 L) & \cos(\lambda_2 L) & \sin(\lambda_2 L) & \cosh(\lambda_3 L) & \sinh(\lambda_3 L) & \cos(\lambda_4 L) & \sin(\lambda_4 L) \\ \sinh(\lambda_1 L)\lambda_1 & \cosh(\lambda_1 L)\lambda_1 & -\sin(\lambda_2 L)\lambda_2 & \cos(\lambda_2 L)\lambda_2 & \sinh(\lambda_3 L) & \cosh(\lambda_3 L) & -\sin(\lambda_4 L)\lambda_4 & \cos(\lambda_4 L)\lambda_4 \\ \cosh(\lambda_1 L)\lambda_1^2 & \sinh(\lambda_1 L)\lambda_1^2 & -\cos(\lambda_2 L)\lambda_2^2 & -\sin(\lambda_2 L)\lambda_2^2 & \cosh(\lambda_3 L)\lambda_3^2 & \sinh(\lambda_3 L)\lambda_3^2 & -\cos(\lambda_4 L)\lambda_4^2 & \sin(\lambda_4 L)\lambda_4^2 \\ \sinh(\lambda_1 L)\lambda_1^3 & \cosh(\lambda_1 L)\lambda_1^3 & \sin(\lambda_2 L)\lambda_2^3 & -\cos(\lambda_2 L)\lambda_2^3 & \sinh(\lambda_3 L)\lambda_3^3 & \cosh(\lambda_3 L)\lambda_3^3 & \sin(\lambda_4 L)\lambda_4^3 & -\cos(\lambda_4 L)\lambda_4^3 \end{bmatrix}$$

$A_L[9...16, 9...16] =$

$$\begin{bmatrix} \cosh(\lambda_1 L) & \sinh(\lambda_1 L) & \cos(\lambda_2 L) & \sin(\lambda_1 L) & -\cosh(\lambda_3 L) & -\sinh(\lambda_1 L) & -\cos(\lambda_4 L) & -\sin(\lambda_1 L) \\ \sinh(\lambda_1 L)\lambda_1 & \cosh(\lambda_1 L)\lambda_1 & -\sin(\lambda_2 L)\lambda_2 & \cos(\lambda_2 L)\lambda_2 & -\sinh(\lambda_3 L)\lambda_3 & \cosh(\lambda_3 L)\lambda_3 & -\sin(\lambda_4 L)\lambda_4 & \cos(\lambda_4 L)\lambda_4 \\ \cosh(\lambda_1 L)\lambda_1^2 & \sinh(\lambda_1 L)\lambda_1^2 & -\cos(\lambda_2 L)\lambda_2^2 & -\sin(\lambda_2 L)\lambda_2^2 & \cosh(\lambda_3 L)\lambda_3^2 & \sinh(\lambda_3 L)\lambda_3^2 & -\cos(\lambda_4 L)\lambda_4^2 & -\sin(\lambda_4 L)\lambda_4^2 \\ \sinh(\lambda_1 L)\lambda_1^3 & \cosh(\lambda_1 L)\lambda_1^3 & \sin(\lambda_2 L)\lambda_2^3 & -\cos(\lambda_2 L)\lambda_2^3 & \sinh(\lambda_3 L)\lambda_3^3 & \cosh(\lambda_3 L)\lambda_3^3 & \sin(\lambda_4 L)\lambda_4^3 & -\cos(\lambda_4 L)\lambda_4^3 \\ 0 & 0 & 0 & 0 & 0 & 0 & 0 & 0 \\ 0 & 0 & 0 & 0 & 0 & 0 & 0 & 0 \\ 0 & 0 & 0 & 0 & 0 & 0 & 0 & 0 \\ 0 & 0 & 0 & 0 & 0 & 0 & 0 & 0 \end{bmatrix}$$